# Electrically Tunable Reflective Metasurfaces with Continuous and Full Phase Modulation for High-efficiency Wavefront Control at Visible Frequencies

*Parikshit Moitra\*, Xuewu Xu, Rasna Maruthiyodan Veetil, Xinan Liang, Tobias W. W. Mass, Arseniy I. Kuznetsov\* and Ramón Paniagua-Domínguez\**

*Institute of Materials Research and Engineering (IMRE), Agency for Science, Technology and Research (A\*STAR), 2 Fusionopolis Way, Innovis #08-03, Singapore 138634, Republic of Singapore*

\*Email: moitra_parikshit@imre.a-star.edu.sg

\*Email: arseniy_kuznetsov@imre.a-star.edu.sg

\*Email: ramon_paniagua@imre.a-star.edu.sg

## Abstract

All-dielectric optical metasurfaces can locally control the amplitude and phase of light at the nanoscale, enabling arbitrary wavefront shaping. However, lack of post-fabrication tunability has limited the true potential of metasurfaces for many applications. Here, we utilize a thin liquid crystal (LC) layer as a tunable medium surrounding the metasurface to achieve a phase-only spatial light modulator (SLM) with high reflection in the visible frequency, exhibiting an active and continuous resonance tuning with associated $2\pi$ phase control and uncoupled amplitude. Dynamic wavefront shaping is demonstrated by programming 96 individually addressable electrodes with a small pixel pitch of ~1 μm. The small pixel size is facilitated by the reduced LC thickness, strongly suppressing crosstalk among pixels. This device is used to demonstrate dynamic beam steering with wide field-of-view and high absolute diffraction efficiencies. We believe that our demonstration may pave the way towards realizing next generation, high-resolution SLMs, with wide applications in dynamic holography, tunable optics and light detection and ranging (LiDAR), to mention a few.

## Introduction

Metasurfaces are one of the fastest growing technologies in nanophotonics. These are ultra-thin planar arrays of nanostructures enabling sub-wavelength control over wavefronts, by changing amplitude, phase and polarization of propagating light, through their spatially varying response.[1–10] In recent years, active tuning of metasurface resonances has opened the door to practical realization of active nanophotonic devices such as dynamic metalenses[11–16], beam steering devices[17–19], tunable holograms[20–22] and many more. Among different tunability approaches[23,24], electrical tuning is the most attractive from the point of view of practical devices. It can be achieved with transparent conducting oxides[18,25–27], two-dimensional materials[28–32], LCs[17,20,33–36] and metallic polymers[37,38], to mention some. LC based metasurface devices, in particular, have made their marks with quite a few important demonstrations with binary and grayscale tuning of LC orientations[17,21,22,39]. A true and continuous grayscale tuning is more interesting as it offers a truly tunable device with multiple phase levels, as demonstrated by LC-tunable SLM based on Fabry-Perot nanocavity[40]. Metasurface-based SLMs[17,41], in their transmissive configuration, however, require thicker LC cell, and have been unable to offer continuous phase gradient as a function of voltage, in turn limiting their functionalities.

Here, we propose a solution to achieve continuous tunable phase gradient using a reflective design with bottom metal layer (which also serves as the pixel electrodes), a dielectric spacer layer and a dielectric metasurface, comprising the SLM device, tunable by LC integration. With in-house nanofabrication, we realize 96 linear aluminum (Al) pixels with small pixel pitch of ~ 1μm, each of which is electrically addressable, to tune the resonant phase of individual metasurface pixel. Continuous phase sweeping through 0-2π, as a function of voltages in a single pixel gives us the power to program the SLM at will to generate any arbitrary wavefront. Using thin LC cell with small pixel sizes, with the flexibility of achieving any arbitrary phase at any pixel location, together with optimized programming of applied voltages, allows us to demonstrate dynamic beam steering with wide field-of-view (FOV) and high efficiencies with low-voltage requirements.

## Results

### Design methodology

Our design rationale aims at realizing a metasurface with high reflection and a complete 2π phase change as a function of orientation of the LC directors in an ultra-thin ($h_{LC} = 500\ nm$) cell configuration. The phase should be modulated when the LC director rotates from its initial state (with the directors oriented in-plane, along the incident light polarization) to the final state (with the directors rotated vertically, along the light propagation direction), with a continuous access to their intermediate rotation states. A metasurface reflect-array design, with the metasurface separated from a bottom metallic reflector with a dielectric spacer, is optimized so that the resonators are in the underdamping regime, which facilitates full 2π phase accumulation at its lowest-order resonance wavelength while being able to keep a high level of reflection[42,43]. Figure 1a shows the schematic of the system, consisting of a planar bottom metallic (Al) reflector

($h_{Al} = 150\ nm$), a silicon dioxide (SiO$_2$) spacer ($h_{Spacer} = 170\ nm$), an array of titanium dioxide (TiO$_2$) disk resonators ($h_{TiO_2} = 200\ nm$, $d_{TiO_2} = 260\ nm$, $P_{TiO_2} = 360\ nm$), an LC cell surrounding the metasurface ($h_{LC} = 500\ nm$), and a top glass superstrate coated by a thin indium tin oxide (ITO, $h_{ITO} = 23\ nm$) layer in a bottom-up configuration, where $h, d$ and $P$ stand for height, diameter and periodicity, respectively. In our finite-difference-time-domain (FDTD, Lumerical) simulation model, when the incident light propagates through the LC cell along the z-axis ($\bar{k} || \hat{z}$) and is polarized along the same direction as the in-plane LC directors (x-axis in Fig. 1a, i.e., $\bar{E} = E_0 \hat{x}$), it "sees" the extraordinary refractive index ($n_e$) of LC. The LC directors can rotate in xz-plane, subtending an angle $\theta_{LC}$ with the z axis (i.e., $\theta_{LC} = 90°$ along x-axis and $\theta_{LC} = 0°$ along z-axis) that can be controlled with an applied bias. At their vertical orientation, the LC directors are said to be switched completely and light "sees" an ordinary refractive index ($n_o$) while passing through the LC cell. We used commercially available LC with $n_e = 1.8131$ and $n_o = 1.522$ (i.e., a birefringence of $\Delta n = 0.291$). In our design, the phase shift of ~$2\pi$ achieved around the metasurface resonance enables full and continuous phase retardation of the reflected wave by tuning the LC director rotation from in-plane to vertical orientation. This can be achieved within a certain range of wavelengths, defined by the initial and final positions of the resonance upon full LC tuning, as seen in Fig. 1b,c. Our resonance of interest is a fundamental metasurface resonance which is centered around $\lambda = 665$ nm wavelength (dashed line in Fig. 1b,c). The reflectance and phase at $\lambda = 665$ nm, as function of LC rotation angle, are shown in Fig. 1d. The figure shows high reflectance and a smooth full $2\pi$ phase sweeping, as a function of LC orientation. We should stress on the fact that this phase modulation is entirely coming from the spectral tuning of the metasurface resonance, and not from the phase accumulation by light propagation in the LC. Indeed, this is easy to corroborate by simply inspecting the phase accumulation upon LC rotation at nearby but off-resonance wavelengths (e.g., <π/2 at $\lambda = 665$ nm, as expected for such a thin LC cell). These results provide the proof-of-concept that our designed metasurface can serve as the core optical component for an SLM device.

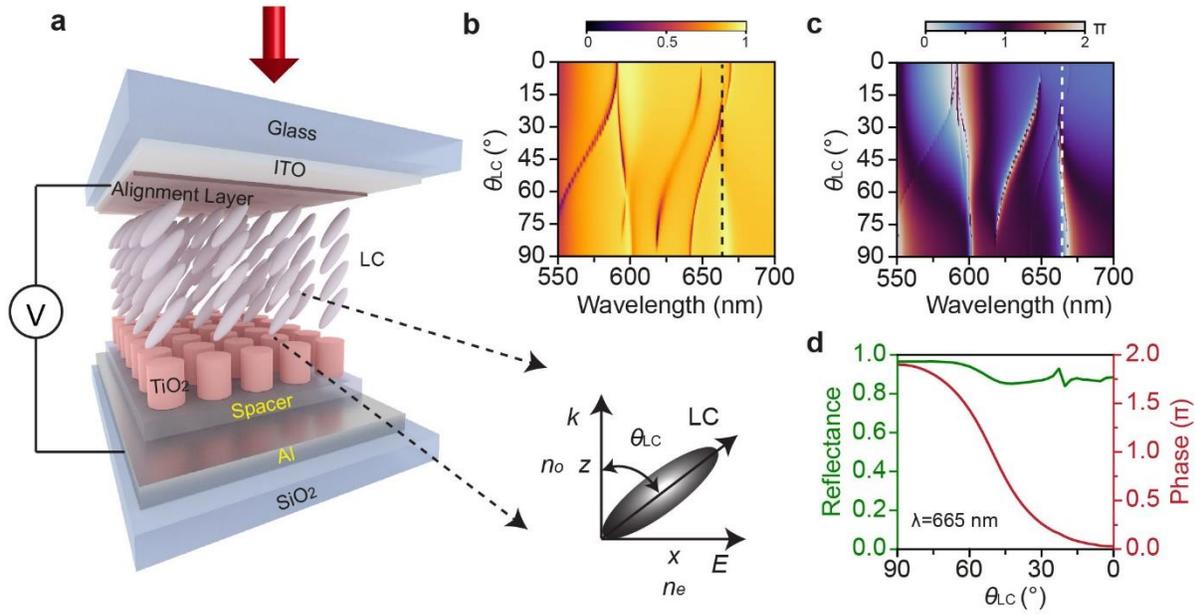

**Fig. 1. Full and continuous phase control using tunable metasurface in reflection. a** Schematic of the tunable metasurface SLM using LC as a tunable medium. Color map of simulated **b** reflectance and **c** phase as a function of wavelength and LC orientation. **d** Reflectance (green) and phase (red) modulation with LC rotation at $\lambda = 665$ nm. Close to 2π continuous phase modulation with high reflectance is demonstrated.

## Tunable metasurface fabrication and packaging

Uncoupling phase modulation from physical propagation in the LC allows the use of an ultra-thin LC cell in our design. This, in turn is expected to lower the voltage requirement for tuning the LC and, at the same time, allow miniaturization of the pixel size by reducing the cross-talk[44], opening the possibility of realizing high-resolution SLMs. To prove this concept, we experimentally realize a metasurface SLM device. The camera image of the final tested device with LC integration on metasurface, chip placement and wire-bonding to a printed circuit board

(PCB) is shown in Fig. 2a. We first fabricate 96 linear Al pixels (electrodes) of 1 μm width ($w_{Al}$) and 1.14 μm pixel pitch ($P_{Al}$), which also serve as the bottom reflector in our design (as shown in the scanning electron microscopy (SEM) images in Fig. 2c). The fan-outs (Al) are fabricated on the same substrate and connected to the electrode pixels (Fig. 2b) with precise alignment, and their ends are wire bonded to the PCB where the substrate is mounted. The metasurface is fabricated on top of a planarized SiO$_2$ spacer layer ($h_{Spacer} = 170\ nm$), separating it from the pixels. After SEM, we corroborate that the nano disk resonators are very close to the design ones, with only a slight tapering resulting in a top diameter of $d_t = 240\ nm$ and bottom diameter of $d_b = 260\ nm$. Figure 2b shows an optical image of the active area of the device with linear Al pixels buried under the metasurface and oxide spacer. For a better visualization of the multiple layers in the device, a focused ion beam (FIB) cross section is carried out at the active area, as shown in Fig. 2d. Due to the invasive nature of milling, the image was taken on a similar device (i.e., from the same batch) as the one tested.

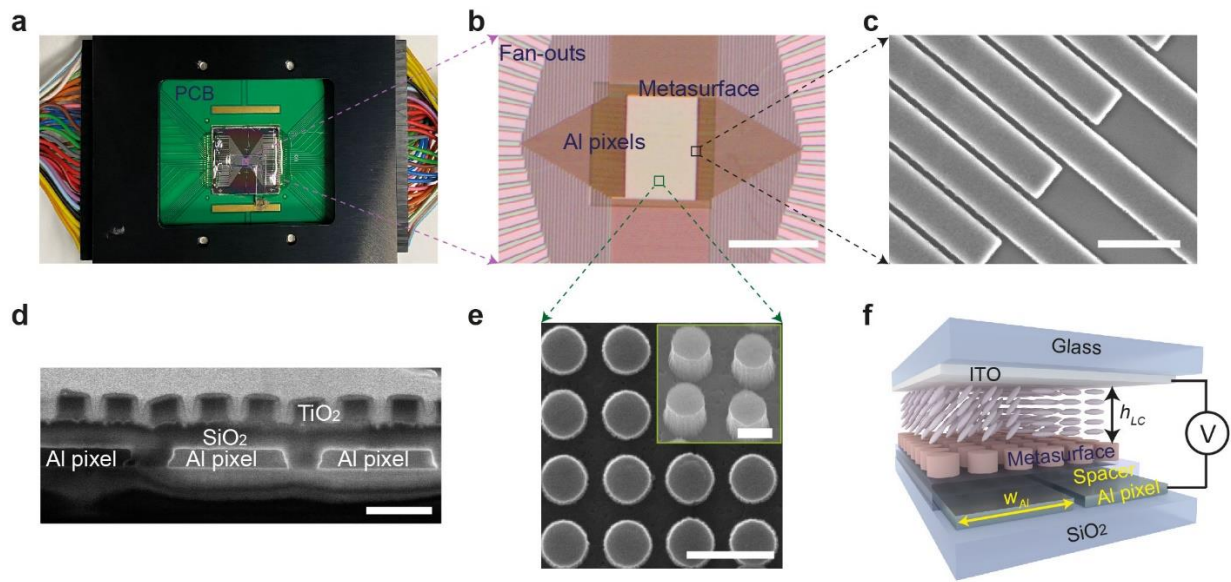

**Fig. 2. Fabrication of tunable metasurface SLM device**. **a** Camera image of the final SLM device and packaging. The substrate is mounted, and wire bonded to a PCB. The 'L'-shaped LC cell and the electrode fan-outs are visible in the image, at the center of which is the active area containing the metasurface (indicated by a purple box). The device is mounted on a customized holder for optical characterization. **b** Optical microscope image of the active area of the device. The Al pixel-array is seen buried under the metasurface. The extensions of the Al pixel-array are connected to the fan-outs. Scale bar is 100 μm **c** SEM image of the Al pixel-array from the array boundary. Scale bar is 2 μm **d** Representative FIB cross section of the active area showing the different layers of the device. A platinum (Pt) coating is applied to protect the nanostructures during milling. Scale bar is 500 nm. **e** SEM image (top view) showing the fabricated metasurface. Scale bar is 500 nm. The inset shows a tilted SEM image (30º). Scale bar is 200 nm. **f** Schematic of the final device.

The SEM images (plan view and tilted) of the metasurface are shown in Fig. 2e before LC infiltration. Finally, LC is infiltrated using capillary forces to the gap created with the aid of a polymide spacer inserted between the substrate and an inverted indium tin oxide (ITO)-glass superstrate. The resulting LC thickness for the tested device is 830 nm, slightly above the design value of 500 nm. The 23 nm ($h_{ITO} = 23\ nm$) thin ITO in the superstate serves as a common electrode and is in direct contact with the LC cell. Fig. 2f schematically shows the SLM architecture. More details on the device fabrication can be found in the Methods section.

## Metasurface resonance tunability

The tunability of the metasurface resonances in our SLM device depends on the orientation of LC directors upon electrical biasing. Hence, the functionality of the device depends on proper addressability of individual electrodes with electrical bias. To check that, we send electrical pulses with a square waveform of $V_{rms}$ = 6 V amplitude at 10 kHz frequency, separately to individual Al pixels and recorded the contrast in reflectance of the biased pixel to the rest of the unbiased pixels in a crossed polarizer-analyzer configuration (more details in Methods section and Supplementary Movie 1). After ensuring that all the pixels are active, we then measure the reflectance spectra of the device as a function of applied voltage, by biasing all the pixels equally. The measured reflectance color map (Fig. 3a) shows the tunability of different resonances with voltage steps of 0.5 V. Comparing to simulations in Fig.1b we can identify the metasurface resonance of interest, centered around $\lambda = 650$ nm, shifting with increasing voltage. The deviation of resonance wavelength from $\lambda = 665$ nm in the design to $\lambda = 650$ nm in the fabricated device stems from the fact that the resonator diameters are slightly smaller in the fabricated device. We note here that the threshold voltage to observe the tunability in the spectrum is as low as 0.5 V and it starts to saturate around 5 V, causing a total spectral shift of ~20 nm. Next, we set up a Michelson interferometer[40] to measure the reflection phase as a function of applied voltage with increments of 0.1 V. The phase colormap (Fig. 3b) shows significant contrast in phase at the resonance wavelength across the voltage range of 0-6 V. We focus on our wavelength of interest ($\lambda = 650$ nm) to better understand the spectrum and phase modulation across the voltage applied, for which we present the reflectance and phase plots versus applied voltage in Fig. 3c. The plots confirm low threshold of ~1 V for phase tuning. The overall phase shift that is experimentally achieved within ~3 V of applied voltage is ~$1.72\pi$. This close to $2\pi$, continuous tunable phase shift opens the opportunity to demonstrate almost arbitrary wavefront shaping. With application of larger voltage, there is some anomalous behavior of phase retardation, which instead of lowering, started to rise to about $0.5\pi$.

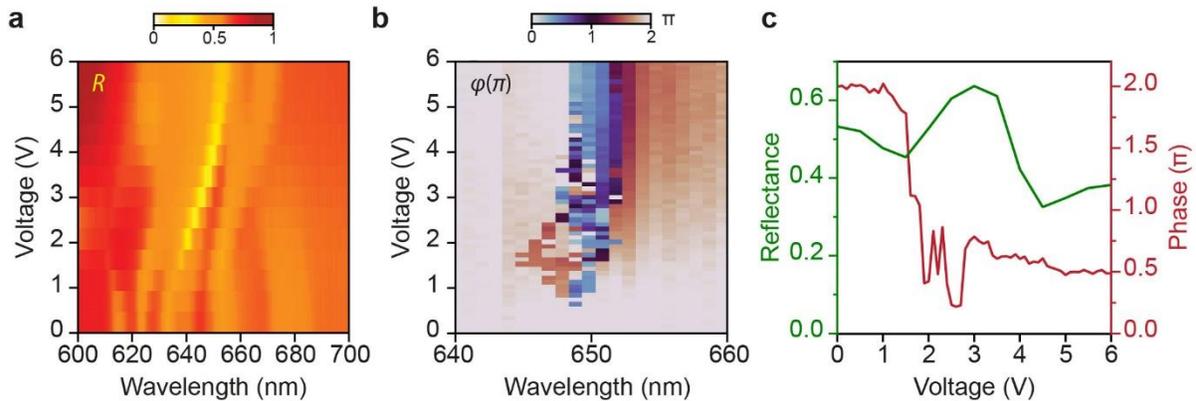

**Fig. 3. Characterization of the tunable metasurface a** Color map of reflectance spectra from the metasurface as a function of applied voltage. Uniform voltage is applied across all 96 electrodes with 0.5 V increment from $V_{rms} = 0$ V

to 6 V.  **b** Phase colormap as a function of applied voltage from $V_{rms}$ = 0 V to 6 V with 0.1 V increment. **c** Tunable reflectance and phase plot at $\lambda$ = 650 nm, versus applied voltage. Close to $2\pi$ phase retardation ($1.72\pi$) is achieved below 3 V with reflectance >50%.

## Dynamic beam steering

Dynamic wavefront shaping with individual pixel control is demonstrated in our device via reconfigurable beam steering. The reason to choose this functionality is two-fold. First, it adapts perfectly to the linear pixel configuration in our metasurface SLM device and, second, it allows precise quantitative measurements of the device efficiency (in turn linked to its reflectivity and ability to finely manipulate the phase in a continuous and full manner). Asserting control over the individual pixels, we created supercells of varying number of pixels in which a linear phase gradient (as close to 0-$2\pi$ as the device allows) is inscribed. Starting from the first pixel, this phase gradient (voltage profile), or supercell, is repeated across the device. Here, we program different supercells containing multiples of 3 pixels (P) i.e., 3P, 6P, 9P and 12P. Once the voltage profile is applied, the linear phase gradient should channel light into the 1$^{st}$ diffraction order, causing beam steering. The direction of beam steering i.e., whether it goes into the +1 or -1 diffraction order, depends on the positive or negative slope of the phase gradient applied, while the steering angle depends on the size of the supercell. Figure 4a schematically illustrates the beam steering mechanism with a 3P supercell configuration, achieved with a 3-level voltage profile ($V_1$, $V_2$ and $V_3$). Without application of voltage gradient, there is no beam steering, as shown in the measured results of Fig. 4b, obtained using a spectrally resolved back-focal plane imaging setup[40].

We optimized the beam steering efficiencies (defined as the power into the desired diffraction order with respect to the incident power) by programming the voltage profiles with a numerical optimization method (see Methods for more details). The optimized voltage profiles for forward bias (corresponding to beam steering into the -1 order) and reverse bias (steering into +1 order) for 3P supercell configurations are given in Fig. 4c. Figure 4d-g show, as color maps, the measured angle-resolved reflection spectrum for increasing number of pixels in the supercells, from left to right. As seen there, the maximum FOV (defined as the diffraction angle multiplied by two) achieved by 3P configuration is ~22º, gradually narrowing for larger supercell sizes, e.g., FOV is ~10º for 6P, ~8º for 9P and ~5º for 12P configurations. The comparison between the diffraction efficiency into the desired -1 order and the +1 order are shown in Fig. 4h-k for increasing supercell sizes. For the smallest supercell, comprising only 3 pixels (3P configuration), we achieved a maximum of ~23% beam steering efficiency at $\lambda = 652$ nm. The efficiency grows significantly for supercells comprising more pixels (as expected from a better phase mapping), reaching up to ~42% for the 6P configuration and nearly 50% for the larger, 9P and 12P supercells. By contrast, the efficiencies in the undesired (+1) diffraction channel stay significantly lower for all these different supercell configurations.

To further demonstrate the reconfigurability of the device, we reversed the phase gradient in the beam steering configuration, in turn reversing the dominating diffraction order from -1 to +1. The results follow closely with the cases with forward biases, showing a maximum FOV of ~22º for 3P configuration and narrowing of the same with larger supercells (Fig. 4l-o). The diffraction efficiencies show analogous trends as well, gradually increasing through 14%, 15%, 25% to a maximum of ~37% at $\lambda = 652$ nm with increasing supercell sizes (Fig. 4p-s). Here we should note that the voltage profile is separately optimized in case of reverse bias. Hence, it does not replicate the same voltage profile as the forward bias. The reason being the behavior of LCs

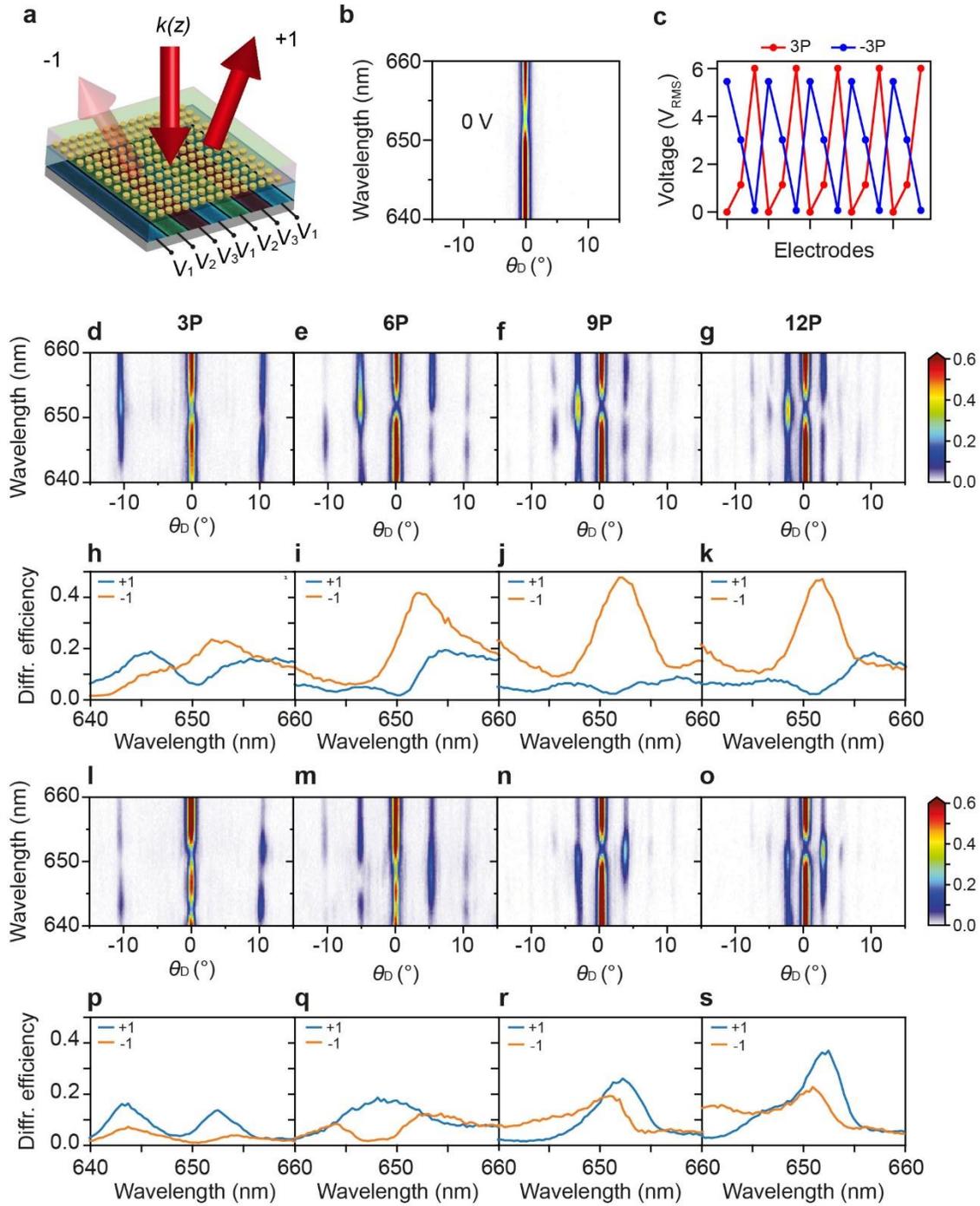

**Fig. 4. Dynamic beam steering with the metasurface SLM a** Schematic of beam steering with 3P supercell configuration with applied voltage profile V1, V2, V3. The light is incident from top and diffracted preferentially into the +1 order. **b** Color map of angle resolved spectra when no bias is applied (0 V). **c** Optimized voltage profiles for 3P configurations for forward and reverse biases. **d-g** Color map of angle resolved spectra for 3P, 6P, 9P and 12P

configurations (forward bias). **h-k** Diffraction efficiency, as extracted from d-g, into the +1 and -1 diffraction order. **l-o** Same as d-g, but in reverse bias. **p-s** Same as h-k, but in reverse bias.

is not as uniform when integrated with metasurfaces with high-resolution nano-structuring, compared with a much simpler liquid crystal on silicon (LCoS) devices, where LCs are integrated on comparatively flat surfaces. Nanofabrication also brings about inhomogeneity and preferential local anchoring of LC molecules inside the metasurface. Moreover, the interpixel crosstalk cannot be ruled out completely.

## Discussion

In conclusion, we have demonstrated a tunable metasurface-based high-resolution reflective SLM device. The metasurface is designed to achieve $2\pi$ phase shift with high reflection at its lower-order resonance. Tunability of the resonance was achieved by incorporating an ultra-thin LC cell as a tunable medium surrounding the metasurface. While LCs help to tune the resonance, and thus the associated phase, upon rotation of their orientations by electrical bias, its ultra-thin character minimizes possible inter-pixel crosstalk. A tunable continuous and full-phase control (close to $2\pi$) is achieved with high reflection at wavelength around $\lambda = 650$ nm. We fabricated the SLM with small pixel size (~1 μm), which is one of the smallest pixel sizes demonstrated by SLM devices. We programmed individual pixels with electrical bias at will and created optimized voltage profiles to encompass a full phase accumulation across a supercell, which is then repeated across the pixels. With this programmable SLM device we achieved tunable beam steering over a wide FOV and with absolute efficiencies up to 50%, which are among the highest reported so far for this class of devices. This demonstration of tunable metasurface paves the way towards realizing novel optical devices for wavefront shaping.

## Methods

**Device Fabrication**

150 nm Al is first sputtered on an 8-inch silicon (Si) wafer with 1 μm thermal oxide, using physical vapor deposition tool (AMAT Endura). The wafer is then diced to 20 mm × 20 mm substrates. To create Al pixels, first, electron beam lithography (Elionix ELS-7000) is carried out using ZEP-520A-7, which is a high-resolution positive tone electron beam resist, followed by developing the exposed pattern using o-xylene developer and rinsing with deionized (DI) water. The patterned electron beam resist serves as a mask for subsequent ICP (inductively coupled plasma) etching (Oxford OIPT Plasma Lab) of the Al film using chlorine ($Cl_2$) and boron trichloride ($BCl_3$) recipe. The remaining resist mask is removed using oxygen ($O_2$) plasma etching. Next, the fan-outs are patterned and aligned with the Al pixel patterns using photolithography (EVG 62008 Infinity) followed by wet etching of Al. The photoresist (S1811) mask is removed by rinsing in acetone. Next, 170 nm of $SiO_2$ spacer is deposited using plasma enhanced chemical vapor deposition (Oxford). The top surface of $SiO_2$ is planarized by the combination of process of spin coating a thick (600 nm) flowable oxide (FOX24), followed by ICP etching using $CHF_3$ recipe to the target thickness of 170 nm. 200nm thick $TiO_2$ is then deposited using ion assisted deposition (Oxford Optofab3000), which is then evaporated (Angstrom) with a 30 nm chromium (Cr) layer and then is subject to electron beam lithography using a high-resolution negative tone resist HSQ (Hydrogen Silsesquioxane) by aligning the metasurface pattern on top of the Al pixel pattern. The resist pattern is developed using a salty developer (NaOH, NaCl)[45], which then serves as the mask for Cr etching using $Cl_2$ and $O_2$ recipe (Oxford OIPT Plasma Lab). $TiO_2$ layer is then etched using left-over HSQ and patterned Cr mask using $CHF_3$ recipe in Advanced Dielectric Etching Systems (SPTS, APS model). The HSQ is totally removed during the etching and the rest of the Cr mask is then removed by wet etching. After fabrication of the metasurface, the substrate is mounted, and wire bonded to a PCB with 1 mil (1 mil = 0.001 inch) gold wires.

Next, the LC cell is integrated into the device. To do that, first a polymide layer is spun on an 'L'-shaped ITO-glass top substrate/superstrate and is cured on a hot plate for 30 minutes at 150 °C. The 'L'-shape was laser-cut from a square superstrate. After that, preferred LC alignment direction is created by unidirectional rubbing. A commercial rubbing machine (Holmarc Opto-Mechatronics - HO-IAD-BTR-03) is employed to control the strength of rubbing. The superstrate is then pressed onto the bottom substrate, with the help of a homemade press, containing multi-point optical profilometer. The 'L'-shape of the superstrate helps to avoid a rough area of the substrate, which may have caused a thicker LC cell, otherwise. The thickness of the cell is determined using the spectrum measured with a portable spectrometer. We then infiltrate commercially available LC QYPDLC-001C (Qingdao QY Liquid Crystal Co. Ltd.) into the cell, and with that the final metasurface SLM device is ready for testing.

**Pixel addressability check**

To set voltages for 96 electrodes we used a 96-channel, 12-bit digital-to-analog converter (DAC) chip on an evaluation board (DAC60096EVM) by Texas Instruments (TI). The DAC board can be addressed by an ESP 32 microcontroller (MCU) via serial peripheral interface (SPI). To facilitate setting and optimization of voltage patterns using python programming, the MCU is connected to a CPU via "universal asynchronous receiver-transmitter" (UART).

To check, if individual pixel can be addressed, we send electrical pulses with square waveform of 6 Vrms amplitude at 10 kHz frequency, separately to individual Al pixels and recorded the contrast in reflectance from the biased pixel from the rest of the unbiased pixels by inserting a set of polarizer and analyzer, oriented mutually perpendicular to each other, with LC director being oriented 45º to the polarizer. The dark contrast in reflection corroborates proper addressability and switching as we expect no cross-polarized light component from the completely switched device with vertically oriented LC directors. The dynamic addressing of individual Al pixels one by one in this configuration is shown in Supplementary Movie 1.

**Optimization of Device Performance**

To optimize beam steering efficiencies, we used the python Application Programming Interface (API) of the NLopt toolbox to optimize the voltage supercells. Our setup enables us to complete a spectral measurement cycle in approximately 0.47 seconds. This process includes an exposure time of 0.25 seconds to achieve a sufficient signal-to-noise ratio, and an overhead time of 0.22 seconds for CCD readout and other processing. Our real-time optimization's cost function evaluation time has a lower limit set by the cycle time. During this interval, a spectral measurement is taken at the back focal plane (BFP), followed by an evaluation of the data and an update of the voltages. Our cost function aims to maximize the light intensity in a specified wavelength and bending angle range, normalized by the total intensity captured in the same wavelength range but from all solid angles as recorded by the objective. In the optimization of phase values, the parameter space forms a hypersphere, however, when considering 96 voltage values with bounds ranging from 0 to 10 V, the parameter space becomes a hypercube. To handle the abrupt voltage changes required for creating phase jumps of 0 to $2\pi$ at the supercell boundaries, we utilized the "Locally biased Dividing RECTangles" (DIRECT-L) algorithm, a derivative-free global optimization method. The algorithm is most effective when applied to parameter spaces with low aspect ratios, like a hypercube. It starts by setting all voltages to the median value of 5 V and uses a large initial step size of 5 V to cover the entire parameter space in the initial divisions. With supercell sizes ranging from 3 to 12 electrodes, the optimization process takes anywhere from 5-

20 minutes, equivalent to a minimum of 300 to 1200 function evaluations respectively. Then, a local optimization, using the "Constrained Optimization By Linear Approximation" (COBYLA) algorithm, is performed to refine the results.

## Acknowledgements

This work was supported in part by the AME Programmatic Grant, Singapore, under Grant A18A7b0058; in part by the IET A F Harvey Engineering Research Prize 2016; and in part by the National Research Foundation of Singapore under Grant NRF-NRFI2017-01. We acknowledge Shi-Qiang Li for his help with initial device driving set up. We acknowledge Norhanani Jaafar for her help with SLM-PCB packaging.

## Conflict of Interest

The authors declare no competing financial interest.

## Contributions

P.M. fabricated the metasurface device, performed initial simulation and optical characterization, and wrote the first draft of the manuscript. X.X. carried out detailed simulation and optimization of the metasurface design and performed electro-optical characterization along with T.W.W.M, who also implemented and performed the beam steering optimization. R.M.V. performed liquid crystal cell fabrication. L.X. performed phase measurements. A.I.K. and R.P.-D. conceived the idea and supervised the work. All authors analyzed the data and contributed to the manuscript preparation.